\documentclass[useAMS,usenatbib]{mn2e}
\usepackage{epsfig}
\usepackage{amsmath}

\title[Simulations on High-$z$ Long GRB Rate]{Simulations on High-$z$ Long Gamma-Ray Burst Rate}
\author[Qin et al.]{Shu-Fu Qin$^{1}$, En-Wei Liang$^{1}$\thanks{E-mail:
lew@gxu.edu.cn}, Rui-Jing Lu$^{1}$, Jian-Yan Wei$^{2}$, Shuang-Nan Zhang$^{3,4}$\\
$^1$Department of Physics, Guangxi University, Nanning 530004, China\\
$^2$National Astronomical Observatories, Chinese Academy of Sciences, 100012
Beijing, China \\
$^3$Key Laboratory of Particle Astrophysics, Institute of High Energy Physics,
Chinese Academy of Sciences, Beijing 100049, China \\
$^4$Physics Department, University of Alabama in Huntsville, Huntsville, AL
35899, USA}
\begin{document}


\pagerange{\pageref{firstpage}--\pageref{lastpage}} \pubyear{2002}

\maketitle

\label{firstpage}

\begin{abstract}
Since the launch of {\em Swift} satellite, the detections of high-$z$ ($z>4$)
long gamma-ray bursts (LGRBs) have been rapidly growing, even approaching the
very early Universe (the record holder currently is $z$=8.3). The observed
high-$z$ LGRB rate shows significant excess over that estimated from the star
formation history. We investigate what may be responsible for this high
productivity of GRBs at high-$z$ through Monte Carlo simulations, with
effective {\em Swif}/BAT trigger and redshift detection probabilities based on
current {\em Swift}/BAT sample and {\em CGRO}/BATSE LGRB sample. We compare our
simulations to the {\em Swift} observations via $\log N-\log P$, peak
luminosity ($L$) and redshift distributions. In the case that LGRB rate is
purely proportional to the star formation rate (SFR), our simulations poorly
reproduce the LGRB rate at $z>4$, although the simulated $\log N-\log P$
distribution is in good agreement with the observed one. Assuming that the
excess of high-$z$ GRB rate is due to the cosmic metallicity evolution or
unknown LGRB rate increase parameterized as $(1+z)^{\delta}$, we find that
although the two scenarios alone can improve the consistency between our
simulations and observations, incorporation of them gives much better
consistency. We get $0.2<\epsilon<0.6$ and $\delta<0.6$, where $\epsilon$ is
the metallicity threshold for the production of LGRBs. The best consistency is
obtained with a parameter set ($\epsilon$, $\delta$)=($\sim 0.4$, $\sim 0.4$),
and BAT might trigger a few LGRBs at $z\simeq 14$. With increasing detections
of GRBs at $z>4$ ($\sim 15\%$ of GRBs in current {\em Swift} LGRB sample based
on our simulations), a window for very early Universe is opening by {\em Swift}
and up-coming SVOM missions.
\end{abstract}


\section{Introduction}
Gamma-ray bursts (GRBs) are the most luminous events in the Universe. They may
be detected out to distance of $z>10$ (Lamb \& Reichart 2000; Bromm \& Leob
2002; Gou et al. 2004; Lin et al. 2004). It is believed that long GRBs (LGRBs)
with $T_{90}>2$ s (Kouveliotou et al. 1993) are from the death of massive stars
(e.g., Woosley 1993; Paczynski 1998; Woosley \& Bloom 2006), hence their rate
may trace the star formation rate (SFR) (Totani 1997; Wijers et al. 1998; Lamb
\& Reichart 2000; Blain \& Natarajan 2000; Porciani \& Madau 2001; Zhang \&
M\'{e}sz\'{a}ros 2004; Piran 2004; Zhang 2007). Simulations for this scenario
can roughly reproduce the observed luminosity ($L$), redshift, and $\log N-\log
P$ distributions (e.g., Liang et al. 2007; Virgil et al. 2009a,b), however,
evidence that LGRBs do not follow star formation unbiasedly has been growing
with rapidly increasing number of high-$z$ GRBs detected with {\em Swift}/BAT
(e.g., Daigne et al. 2006; Le \& Dermer 2007; Y\"uksel \& Kistler 2007;
Salvaterra \& Chincarini 2007; Guetta \& Piran 2007; Salvaterra et al. 2008;
Kistler et al. 2008, 2009; Wang \& Dai 2009). After the detection of GRB 050904
at $z=6.29$ (Kawai et al. 2006; Haislip et al. 2006) three year later, other
two higher-$z$ LGRBs were detected, GRB 080913 at $z\sim6.7$ (Greiner et al.
2009) and GRB 090423 at $z\sim 8.3$ (Tanvir et al. 2009). This ever-increasing
number of detectable high-$z$ LGRBs shows that the LGRB rate at high redshift
may be higher than that expected previously. Kistler et al. (2008) found
significant excess of the LGRB rate over the SFR at $z\approx 4$ based on a
sample of 36 luminous LGRBs. They suggested that some unknown mechanisms lead
to this observed enhance in the GRB rate, such as the cosmic evolutions of the
metallicity history and LGRB luminosity function.

Cosmic chemical evolution predicts that LGRBs should occur preferentially in
metal poor environment. From a theoretical point of view, this is not
surprising since lower metallicity leads to weaker stellar winds and hence less
angular momentum loss, more probably resulting in rapidly rotating cores in the
stars, as required in the collapsar model for GRBs (e.g., Woosley 1993;
MacFayden \& Woosley 1999; Yoon \& Langer 2005). With host galaxy observations
(Fynbo et al. 2003; Conselice et al. 2005; Gorosabel et al. 2005; Chen et al.
2005; Starling et al. 2005; Fruchter et al. 2006; Fynbo et al. 2006; Mirabal et
al. 2006), it is found that the metallicity of LGRB hosts are on average lower
than that of core-collapse supernovae (Wolf \& Podsiadlowski 2007). The {\em
Swift} sample also suggests a modest propensity for low-metallicity, evidenced
by an increase in the rate density with redshift (Butler et al. 2009). One
major restriction of LGRBs to low metallicity is that LGRBs do not follow star
formation in an unbiased manner and LGRB rate may peak at a significantly
higher redshift than supernovae (Lin et al. 2004; Langer \& Norman 2006;
Firmani et al. 2006). Assuming that LGRB rate traces both the SFR and the
metallicity evolution history, Li (2008) found that the observed redshift
distribution of the {\em Swift} LGRBs can be reproduced with a fairly good
accuracy.

It was also suggested that LGRBs may have experienced some sort of luminosity
evolution with redshift, being more luminous in the past (Lloyd-Ronning et al.
2002; Firmani et al. 2004; Kocevski\& Liang 2006; Salvaterra et al. 2008;
Salvaterra et al. 2009) or having a higher efficiency of LGRB production rate
by massive stars at high-$z$ (Daigne, et al. 2006). This effect may also modify
the observed LGRB rate.

This paper is dedicated to exploring what may be responsible for the high
productivity of LGRBs at high-$z$ through simulations. Four scenarios are
tested: (A) A LGRB rate that purely follows SFR; (B) a LGRB rate that follows
SFR in cooperation with some sort of LGRB rate evolving with redshift
characterized as $(1+z)^{\delta}$ (Kistler et al. 2009); (C) a LGRB rate that
follows SFR in cooperation with cosmic metallicity evolution (taking the form
as Langer and Norman 2006); (D) a LGRB rate that follows the SFR in conjunction
with both LGRB rate evolution and cosmic metallicity evolution. Throughout this
paper a flat Universe with $H_0=71$ km s$^{-1}$Mpc$^{-1}$, $\Omega_M$=0.3 and
$\Omega_\Lambda$=0.7 is assumed.

\section[general]{Model}
The number of LGRBs per unit time at redshift $z\sim z+dz$ with luminosity
$L\sim L+dL$ is given by
\begin{equation}\label{dn}
\frac{d N}{dtdzdL}=\frac{R_{\rm
LGRB}(z)}{1+z}\frac{dV(z)}{dz}\Phi(L),
\end{equation}
where $R_{\rm LGRB}(z)$ is the event rate of LGRB (in units of
${\rm Gpc^{-3} yr^{-1}}$) as a function of $z$, $(1+z)^{-1}$
accounts for the cosmological time dilation, $\Phi(L)$ is the LGRB
luminosity function, and
\begin{equation}\label{dvdz}
\frac{dV}{dz}=\frac{c}{H_0}\frac{4\pi D^2_L}{(1+z)^2
[\Omega_M(1+z)^3+\Omega_\Lambda]^{1/2}}
\end{equation}
is the co-moving volume element at redshift $z$ in a flat $\Lambda$ CDM
universe. The $D_L(z)$ is the luminosity distance at $z$. Considering an
instrument with flux threshold $P_{\rm th}$ and an average solid angle $\Omega$
for the aperture flux, the number of LGRB triggers with peak luminosity in a
range [$L_{\rm min}$, $L_{\rm max}$] during an observational period $T$ should
be
\begin{equation}\label{N}
N=\frac{\Omega T}{4\pi}\int_{L_{\min}}^{L_{\max}}
\eta_t(P)\Phi(L)dL\int_0^{z_{\max}} \frac{R_{\rm
LGRB}(z)}{1+z}\frac{dV(z)}{dz}dz,
\end{equation}
where $\eta_t(P)$ is the trigger probability of a burst with observed peak flux
$P$, given by $P=L/4\pi D_L^2(z)k$, where the $k$ factor corrects the
bolometric flux in the burst rest frame ($1-10^{4}$ keV in this analysis) into
the observed flux in an instrument energy band with the spectral information.
The $z_{\max}$ of a given burst with luminosity $L$ is determined by the
instrumental flux threshold $P_{\rm th}$ through $P_{\rm th}=L/4\pi
D^2_L(z_{\max})k$.

As discussed in \S 1, we express the LGRB rate as a function of
redshift with
\begin{equation} \label{RGRBs}
R _{\rm LGRB}(z) \propto R_{\rm SFR} (z) f(z)\Theta (\epsilon, z),
\end{equation}
where $R_{\rm SFR}(z)$ is the star formation rate, $f(z)$ accounts for
potential cosmological evolution effects, $\Theta (\epsilon, z)$ is the
fractional mass density belonging to metallicity below $\epsilon Z_{\odot}$  at
a given $z$ ($Z_{\odot}$ is the solar metal abundance), and $\epsilon$ is
determined by the metallicity threshold for the production of LGRBs. In our
simulations, $f(z)$ is simplified to $f(z)=(1+z)^{\delta}$ as adopted by
Kistler et al. (2009), and the $\Theta (\epsilon, z)$ follows the law (Langer
\& Norman 2006),
\begin{equation} \label{metallicity}
\Theta (\epsilon, z)= \frac{\hat{\Gamma}(\kappa+2,\epsilon^\beta 10^{0.15\beta
z} )}{\Gamma(\kappa+2)},
\end{equation}
where $\kappa=-1.16$ is the power-law index in the Schechter distribution
function of galaxy stellar masses (Panter et al. 2004), $\beta=2$ is the slope
in the linear bisector fit to the galaxy stellar mass-metallicity relation
(Savaglio 2006), and $\hat{\Gamma}(a, x)$ and $\Gamma(x)$ are the incomplete
and complete gamma function, respectively. The value of $\epsilon$ varies from
$0.2$ to $0.6$ (Modjaz et al. 2008, Li 2008).

The star formation history is not well constrained by data at high redshifts
($z>4$). The current observed SFR rapidly increases at $z<1$, keeps almost a
constant up to $z=4.0$, then drops at $z>4$. The sharp drop at $z>4$ may be due
to large dust extinction at such high redshifts. We adopt a SFR parameterized
as (Rowan-Robinson 1999; Hopkins $\&$ Beacom 2006),
\begin{equation} \label{SFR}
R_{\rm SFR}(z)=R_0
\begin{cases}
(1+z)^{3.44} \quad\quad \quad  \  \ z \leq z_{peak}, \\
(1+z_{peak})^{3.44}\quad \quad  z > z_{peak}, \\
\end{cases}
\end{equation}
where $z_{peak}=1$ and $R_0$ is a normalization parameter. The slope of the
segment $z \leq z_{peak}$ is the well-fitted parameter in Hopkins $\&$ Beacom
(2006).

It was suggested that low luminosity GRBs (LL-GRBs) may be a distinct GRB
population from high-luminosity GRBs (HL-GRBs) with much higher local event
rate (Soderberg et al. 2004; Cobb et al. 2006; Liang et al. 2007; Chapman et
al. 2007). Liang et al. (2007) proposed a two-component broken power-law to
model the GRB luminosity function. They characterize the luminosity function
for each GRB population as
\begin{equation}
\phi(L)=\phi_0[(\frac{L}{L_b})^{\alpha_1} +(\frac{L}{L_b})^{\alpha_2}]^{-1},
\end{equation}
where $\phi_0$ is a normalization constant to assure the integral over the
luminosity function being equal to unity, and $L_b$ is the break luminosity. We
do not consider the cosmological evolution of the luminosity
function\footnote{The luminosity evolution with being luminous in the past
(Lloyd-Ronning et al. 2002; Firmani et al. 2004; Kocevski\& Liang 2006;
Salvaterra et al. 2008; Salvaterra et al. 2009) may also result in an increase
of GRB rate with redshift as $(1+z)^{\delta}$ (Kistler et al. 2008).}. The
global luminosity function thus is given by
\begin{equation}\label{LF}
\Phi(L)=\Phi_0[\rho^{LL}_{0}\phi^{\rm LL}(L)+\rho^{HL}_{0}\phi^{\rm HL}(L)]
\end{equation}
where $\Phi_0$ is a normalization constant, $\rho^{\rm LL}_{0}$ and $\rho^{\rm
HL}_{0}$ are the local rates of low and high luminosity LGRBs, respectively.

\section{{\em Swift} GRB sample and {\em Swift}/BAT trigger probability}

In order to eliminate the selection effect of different instruments, we use
only the LGRBs detected by {\em Swift}/BAT to form a homogeneous sample. As of
31 May 2009, {\em Swift} had detected 385 LGRBs, with redshift measurement for
124 GRBs. The peak fluxes, measurement redshift and the spectral information
for the GRB in our sample are taken from the NASA website\footnote{
http://swift.gsfc.nasa.gov/docs/swift/archive/grb\_table/}. The peak fluxes are
measured on 1-s timescale. It is known that the GRB spectrum is adequately fit
with a smooth broken power-law, the so-called Band function (Band et al. 1993).
Because of the narrowness of the BAT energy band, the BAT data cannot
adequately constrain the spectral parameters of GRBs, and the GRB spectra
observed with BAT are well fitted with a simple power law, $N\propto
\nu^{-\Gamma}$ (Zhang et al. 2007). Empirically, $\Gamma$ is roughly correlated
to the observed peak spectral energy $E_{\rm p}$ by $\log E_{p}=(2.76\pm 0.07)
- (3.61\pm 0.26) \log \Gamma$ (Zhang et al. 2007; Sakamoto et al. 2009). We
employ this empirical relation to estimate the $E_{\rm p}$ of {\em Swift} GRBs
and correct the observed peak luminosity to a bolometric band ($1-10^4$ keV)
assuming the low and high energy spectral indexes as $\Gamma_1=0.83$ and
$\Gamma_2=2.35$ (Preece et al. 2000; Kaneko et al. 2006) for all {\em Swift}
GRBs and simulated bursts.

Selection effects involved in a GRB sample may disguise its truly intrinsic
distribution (Bloom 2003; Fiore et al. 2007; Kistler et al. 2008; Le \& Dermer
2007; Coward 2007; Coward et al. 2008). Two kinds of selection effects are
concerned in our simulations. One is the {\em Swift}/BAT GRB trigger and the
other is the redshift determination through spectroscopy of the optical/NIR
afterglow or of the GRB host galaxy. The first one affects the peak flux
distribution of a GRB sample and the other one affects the $L-z$ distribution.

\begin{figure}
\resizebox{8cm}{!}{\includegraphics{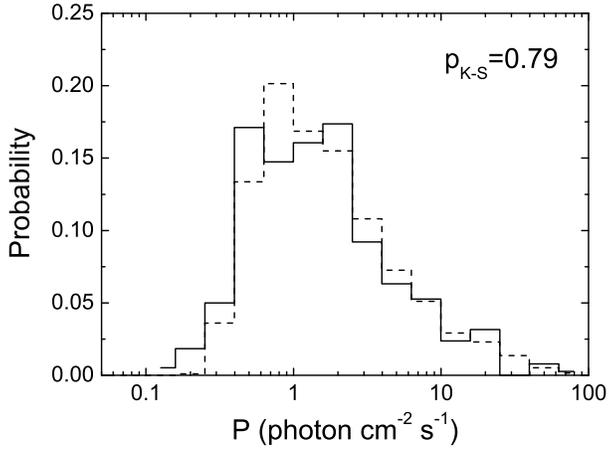}}
\caption{Comparison of the peak
flux distribution of LGRBs detected by BATSE (dashed line) to BAT GRB sample
(solid line). BATSE data are from Stern et al. (2001) and BAT data are from
NASA website. The peak fluxes of the BATSE GRBs are converted to the BAT energy
band of [15, 150] keV for comparison, with a typical spectral parameter set of
$\Gamma_1=0.83, \Gamma_2=2.35$ and $E_{p}=302$ keV. } \label{fPF}
\end{figure}

The {\em Swift}/BAT trigger is quite complex and its sensitivity for GRBs is
very difficult to parameterize (Band 2006). Practically, a GRB with peak flux
slightly over the instrument threshold would not always make a trigger. The BAT
trigger rate is about 64 LGRBs per year per solid angle, roughly equal to that
of BATSE ($\sim 67$ LGRBs per year per solid angle, see Stern et al. 2001).
Comparison of the peak flux distributions of the LGRBs detected with BAT and
BATSE is shown in Fig. \ref{fPF}. The Kolmogorov-Smirnov Test (K-S test) yields
a probability $p_{\rm K-S}=0.79$, indicating that the two distributions are
statistically indistinguishable. Therefore the two detectors should have
comparable sensitivity for LGRBs. We consequently use the same trigger
probability of {\em Swift}/BAT as that of {\em CGRO}/BATSE.

\begin{figure}
\resizebox{8cm}{!}{\includegraphics{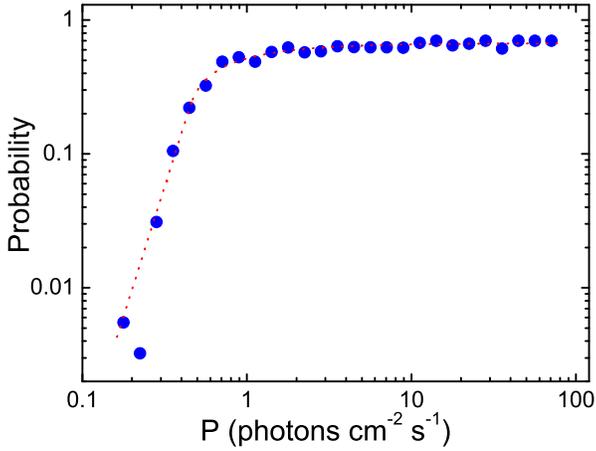}} \caption{Trigger
probability as a function of the peak flux with our best fit (dotted curve) for
BATSE LGRBs. The best fit curve is parameterized in Eq.
\ref{trig_eff}.}\label{fEff}
\end{figure}

Stern et al. (2001) scanned the off-line daily archival data from BATSE and
identified 3713 LGRBs, of which 1916 LGRBs are triggered events. Assuming that
the peak fluxes of both trigged and non-trigged GRBs are over the BATSE
threshold, we define the trigger efficiency $\eta_t$ of a GRB as a function of
$P$ with the ratio of trigger event number to the total detected event number
(including both triggered and non-trigged events) in a peak flux bin $P+dP$.
The result is shown in Fig. \ref{fEff}, where $P$ is converted to {\em Swift}
energy band [15, 150] keV with the observed spectral parameters. It is found
that the instrument trigger threshold is roughly 0.2 photon s$^{-1}$ cm$^{-2}$.
$\eta_t$ as a function of $P$ can be parameterized as
\begin{equation}\label{trig_eff}
\eta_{t}=
\begin{cases}
5.0 P^{3.85} , P<0.45 \\
0.67 (1.0-0.40/P)^{0.52}, P\geq0.45
\end{cases}.
\end{equation}
Note that $\eta_t$ remains almost a constant at $67\%$ for $P>1$ counts
cm$^{-2}$ s$^{-1}$. Some non-triggered events are very bright (the strongest
one is 24 ph cm $^{-2}$ s$^{-1}$). This is due to the dead time when the
trigger was disabled during data readouts or when {\em CGRO} passed through
regions of very high ionospheric activity (Stern et al. 2001). This effect does
not affect our our simulations. We use this trigger probability curve for {\em
Swift}/BAT in our simulations.

\begin{figure}
\resizebox{8.2cm}{!}{\includegraphics{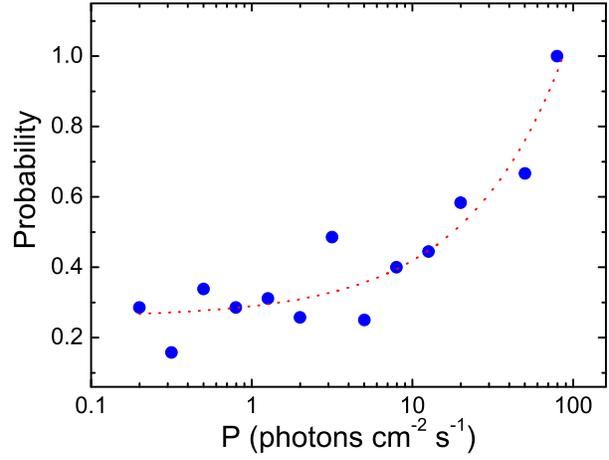}}
\caption{Redshift
measured probability as a function of the peak flux with our best fit (dotted
curve) for BAT LGRBs. The best fit curve is parameterized in Eq.
\ref{z_pro}.}\label{pzpro}
\end{figure}

Fiore et al. (2007) pointed out that about 30\% of {\em Swift} GRBs have
measured redshifts (124/385 in the current {\em Swift} GRB sample). In order to
simulate a GRB sample with redshift measurement, we empirically model
probability of redshift measurement for a BAT triggered burst. From the current
GRB sample, redshifts are preferentially measured for brighter GRBs at lower
redshifts. We show the redshift measurement probability $\eta_z$ as a function
of $\log (P)$ in Fig. \ref{pzpro}, where the probabilities are calculated by
the ratio of the number of GRBs with redshift measurement in each bin to the
number of triggered GRBs in the corresponding bin. It is found that the
probability of redshift measurement for the burst in our sample does not
strongly depend on $P$. In fact, the redshift measurement is complicated,
depending on many artificial effects, such as the optical follow-up, spectral
line detection, etc (Bloom 2003). Even though, we still use an empirical
function parameterized as
\begin{equation}\label{z_pro}
\eta_{z}=0.26+0.032 e^{1.61\log P}
\end{equation}
to simulate GRB samples with redshift measurement.
\section{Monte Carlo Simulations}
Based on the model discussed above, we make simulations to re-produce the {\em
Swift} observations. In our simulations, the model parameters are constrained
with comparison of our simulations to observations. Note that the parameters of
the LL-GRB luminosity function are similar to that reported by Liang et al.
(2007), i.e., $\alpha_1^{LL}=0$, $\alpha_2^{\rm LL}=3.0$, and $L_b^{\rm
LL}=7.5\times10^{46}$ erg s$^{-1}$. For HL-GRBs, we find that $L_b^{\rm
HL}=2.75\times10^{52}$ erg s$^{-1}$ and $\alpha_1^{\rm HL}=1.36$ can well
reproduce the observations at $L<L_b^{\rm HL}$. This is roughly consistent with
that reported by Stern et al. (2002), Schmidt (2001), and Lloyd-Ronning et al.
(2004). We fix these parameters in all cases we study below. The parameter
$\alpha_2^{\rm HL}$ are slightly adjustable, varying from 2.0 to 2.5, in order
to get the best consistency between our simulations and observations for
different cases. The $\rho^{\rm HL}_{0}$ is $\sim 1$ Gpc$^{-3} $ year$^{-1}$
(e.g., Schmidt 2001; Stern et al. 2002; Lloyd-Ronning et al. 2004; Liang et al.
2007), but it was suggested that $\rho^{\rm LL}_{0}=100\sim 1000$ Gpc$^{-3} $
year$^{-1}$, much higher than $\rho^{\rm HL}_{0}$ (e.g., Cobb et al. 2006;
Liang et al. 2007; Chapman et al. 2007). The $\rho^{\rm LL}_{0}$ is very
uncertain, but it does not significantly affect the K-S test results since only
a few low-luminosity GRBs in the simulated GRB sample. The $\epsilon$ and
$\delta$ that describe the metallicity history and the GRB rate evolution,
respectively, are free parameters. We search for the best model parameter sets
by evaluating the consistency between the simulated $P$, $L$, and $z$
distributions and the observed ones with the K-S test. A larger value of
$p_{K-S}$ indicates a better consistency. A value of $p_{K-S}>0.1$ is generally
acceptable to claim the statistical consistency, and a value of
$p_{K-S}<10^{-4}$ convincingly rejects the hypothesis of the consistency (e.g.
Bloom 2003). For simulations for a given set of parameters, we first evaluate
the consistency of the $\log N-\log P$ distribution for the simulated triggered
sample since the $L$ and $z$ distributions suffer from the biases of redshift
detection. By screening the simulated triggered sample with the redshift
measurement probability curve the consistencies of the $P$, $L$ and $z$
distributions should be significantly improved (with $p_{\rm K-S}
>0.1$) for a reasonable parameter set. The details of our procedure are
described as follows.
\begin{enumerate}
\item Simulate a GRB with luminosity $L$ at redshift $z$, GRB$(L, z)$. The
redshift is generated from the co-moving number density of GRBs at redshift
$z+dz$, i.e.,
\begin{equation}
\Re (z)=\frac{R_{\rm LGRB}(z)}{1+z}\frac{dV}{dz},
\end{equation}
and $L$ is simulated with the probability distributions derived from Eq.
\ref{LF}. We take $[L_{\min},\ L_{\max}]=[10^{45},\ 10^{55}$] erg s$^{-1}$ and
a redshift range of $z=0-20$.

\item Derive $E_{\rm p}^{'}$ in the burst frame with the $L-E_{\rm p}^{'}$
relation,
\begin{equation}
E_{p}^{'}\approx 200 {\rm keV}L_{52}^{1/2}/C,
\end{equation}
where $C$ is taken from a random distribution in the range of [0.1, 1](Liang et
al. 2004). The photon indices prior and post the break energy are assumed to be
$0.83$ and $2.35$, respectively. With the spectral information, we calculate
the peak photon flux $P$ in the BAT band that corresponds to an energy band
$[15\times (1+z),\ 150\times (1+z)]$ in the burst frame. The simulated GRB then
is characterized as GRB $(L,z,P)$.

\item Screen a mock GRB with the BAT trigger probability (Eq. \ref{trig_eff}).
We get the trigger probability $\eta_t$ of burst with peak flux $P$ from Eq.
\ref{trig_eff} and generate a random number $Q_t$ in the range (0,1). If
$0<Q_t<\eta_t$, we pick up this event as a triggered GRB.

\item Repeat the above steps to simulate a BAT-triggered GRB sample of 385 GRBs
and evaluate the consistencies of its $\log N-\log P$ distribution with the BAT
observations by the K-S test.

\item Screen the simulated triggered sample (385 GRBs) with the redshift
measurement probability curve (Eq. \ref{z_pro}). To do so, we calculate the
$\eta_z$ value for a given burst, and generate a random number $Q_z$ in the
range (0,1). If $0<Q_z<\eta_z$, we pick up this event as a triggered GRB with
redshift detection.

\item Evaluate the consistencies of $\log N-\log P$, $L$ and $z$ distributions
of the simulated known-redshift by the K-S test. In a reasonable parameter set
these distributions should be greater agreement with the observations than that
of the simulated triggered sample, i.e., with larger $p_{K-S}$ values.
\end{enumerate}

We create simulations for four cases as below. We present our selected
simulation results in the following section. The selected model parameters and
K-S test results are summarized in Table 1, and distributions are shown in
Figs. 4-6 and 8. Our simulations show that the highest redshift of LGRBs that
may be triggered with {\em Swift}/BAT is $\sim 14$.

\subsection{Case A: $R_{\rm LGRB}(z) \propto R_{\rm SFR} (z)$}
\begin{figure}
\resizebox{8cm}{!}{\includegraphics{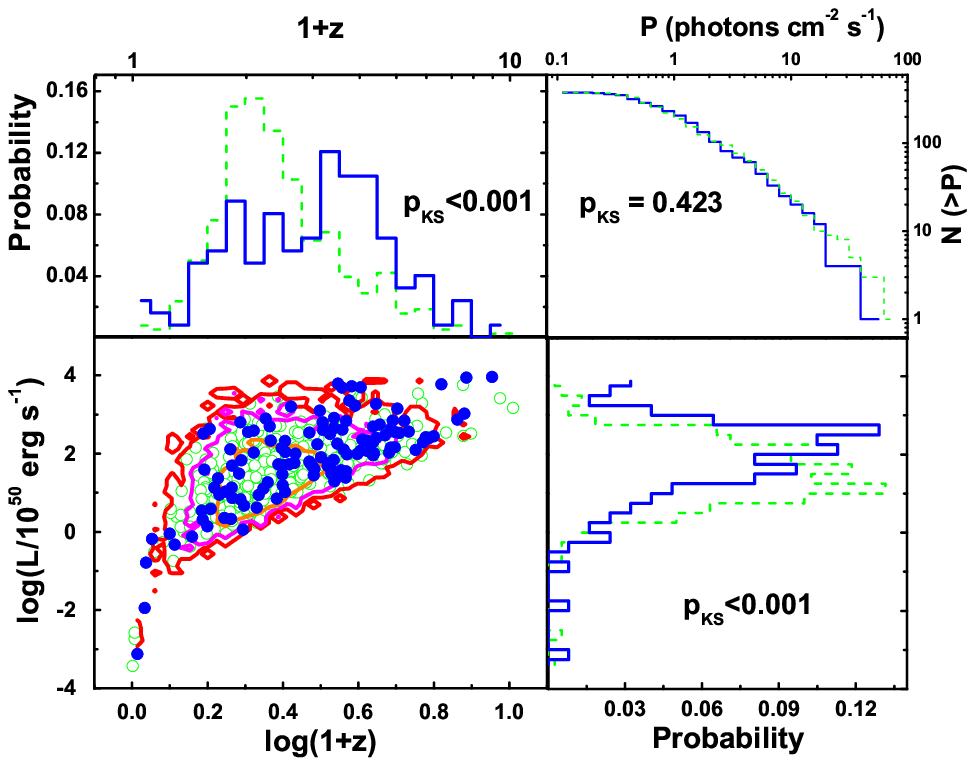}}
\resizebox{8cm}{!}{\includegraphics{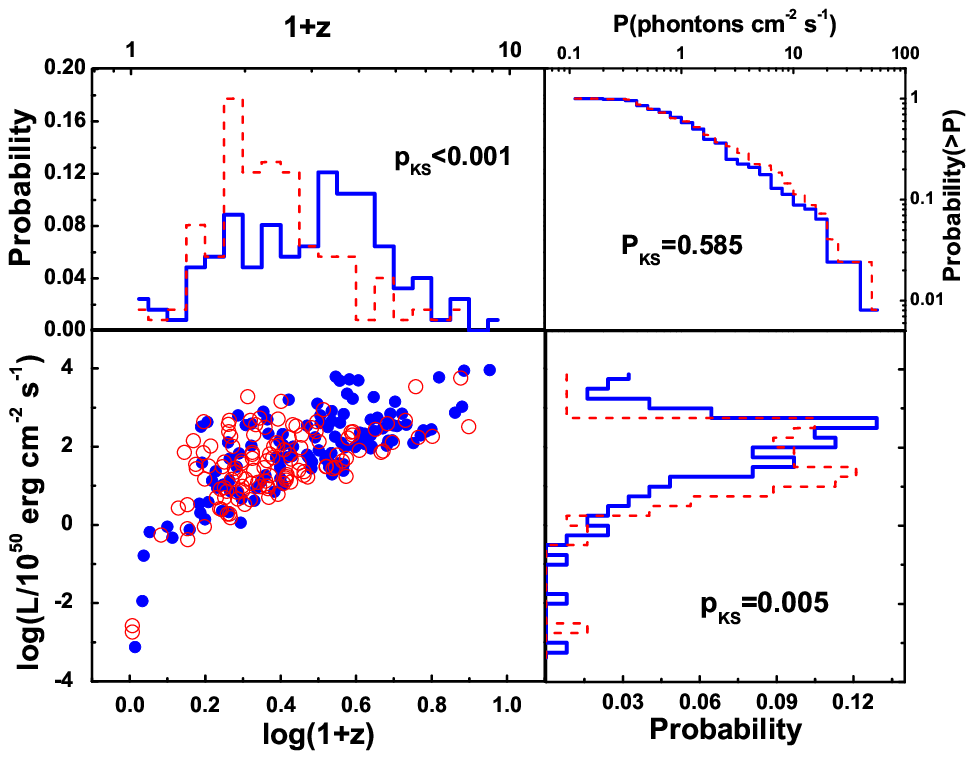}} \caption{Comparisons of
two-dimensional $\log L-\log z$ distributions and one-dimensional $\log L$,
$\log z$, and $\log P$ (accumulative $\log N-\log P$) distributions between the
observed {\em Swift}/BAT GRB sample (solid dots and solid lines) and our
simulations (open dots and dashed lines) for Case A: $R_{\rm LGRB}(z) \propto
R_{\rm SFR} (z)$. {\em left four panels} are for the trigged GRB sample. The
contours in the $\log L-\log z$ plane show the relative probability
distributions of the simulated GRBs with different color lines: red for
$99.7\%$, magenta for $95.5\%$, and orange for $68.7 \% $. {\em Right four
panels} are for the sample with redshift measurement. One dimensional K-S test
probabilities for the comparisons are marked in each panel.} \label{Case_A}
\end{figure}

This case is for the GRB rate that purely follows the SFR. This was performed
by Virgili et al. (2009a). We make improvements in this case by considering the
trigger probability and redshift detection probability in our simulations.
Comparisons between simulations and {\em Swift} observations are shown in Fig.
\ref{Case_A}. It is found that our simulations well reproduce the observed
$\log N-\log P$ distribution, yielding $p_{\rm K-S}=0.423$ and $p_{\rm
K-S}=0.585$ for the trigged and known-redshift samples, respectively. However,
the observed luminosity and redshift distributions are poorly reproduced. The
simulated GRBs at $z\approx 1$ is significantly over-produced, but the GRBs at
$z>2$ have significant deficit. This case also over-produces the GRBs around
$L=10^{51}$ erg s$^{-1}$, and under-generates GRBs with $L>10^{53}$ erg
s$^{-1}$. The $p_{K-S}$ for the $\log L$ and $\log z$ distributions are smaller
than 0.01.

\subsection{Case B: $R_{\rm LGRB}(z) \propto  R_{\rm SFR} (z) \Theta (\epsilon, z)$
} 
\begin{figure}
\resizebox{8cm}{!}{\includegraphics{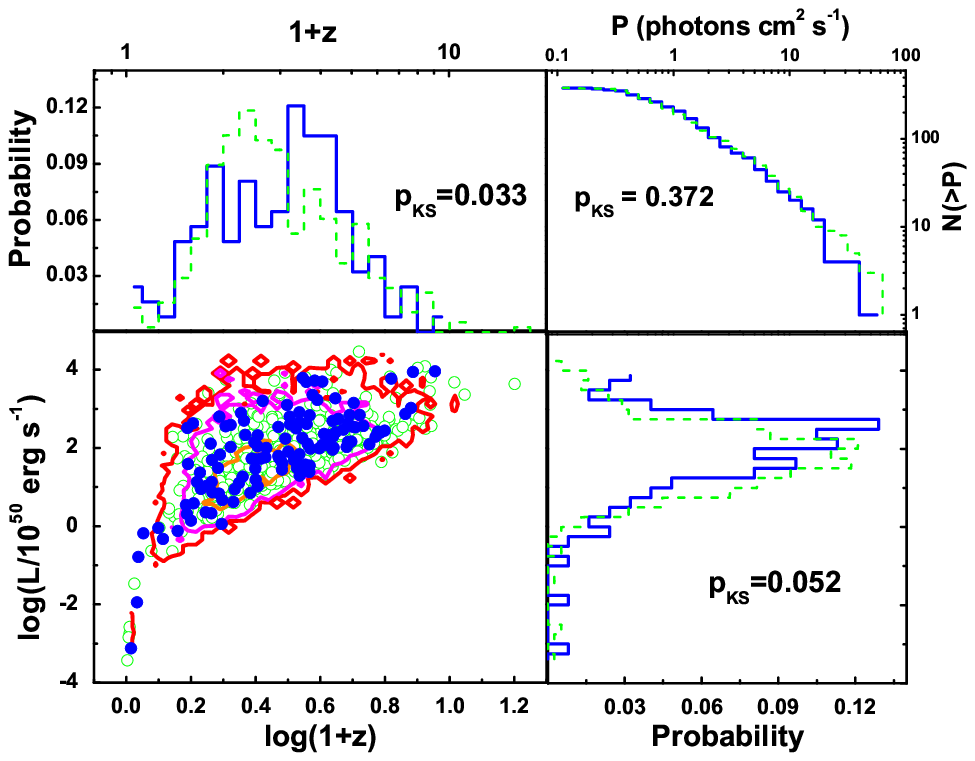}}
\resizebox{8cm}{!}{\includegraphics{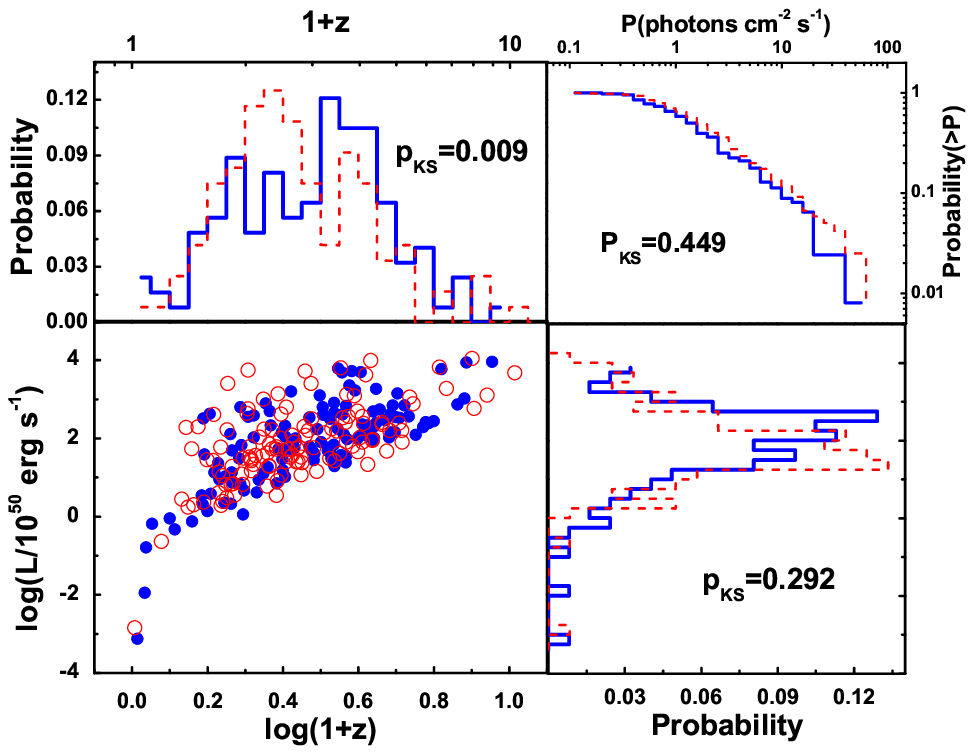}} \caption{The same as Fig. 4, but
for Case B: $R_{\rm LGRB}(z) \propto R_{\rm SFR} (z) \Theta (\epsilon, z)$.}
\label{Case_B}
\end{figure}

This case assumes that the GRB rate is proportional to the star formation
history incorporating with the cosmic metallicity history. We find that
$\epsilon\sim 0.5$ gives the best consistency. Our results are shown in Fig.
\ref{Case_B}. In general, the simulations for this case are more consistent
with the observations than Case A. The $\log N-\log P$ distribution is well
reproduced, with $p_{\rm K-S}=0.372$ and $p_{\rm K-S}=0.449$ for the
simulations on the trigged and known-redshift samples, respectively. The
consistency of the simulated luminosity distribution is also accepted, with
$p_{K-S}=0.052$ and $p_{K-S}=0.292$ for the triggered and the known-redshift
samples, respectively. However, the simulated $z$ distributions are only
marginally acceptable, with $p_{\rm K-S}=0.033$ and $p_{\rm K-S}=0.009$ for the
simulations on the trigged and known-redshift samples, respectively. Note that
the consistency of $z$ distribution becomes much worse for the simulated
known-redshift sample.

\subsection{Case C: $R_{\rm LGRB}(z) \propto R_{\rm SFR} (z) f(z)$}
\begin{figure}
\resizebox{8cm}{!}{\includegraphics{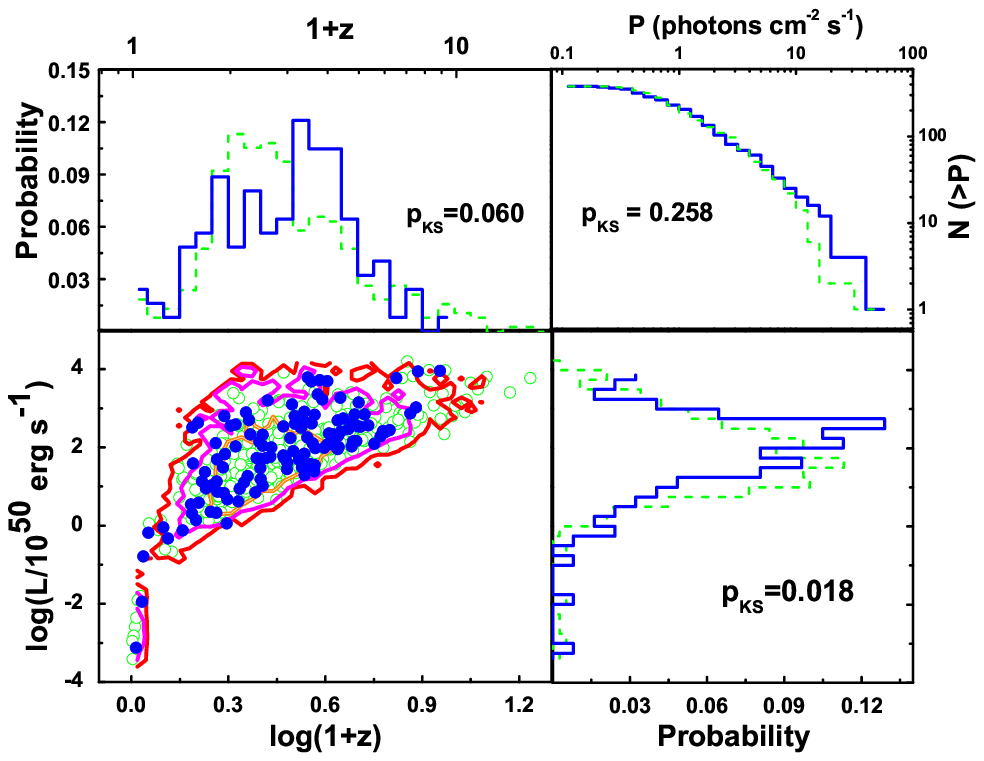}}
\resizebox{8cm}{!}{\includegraphics{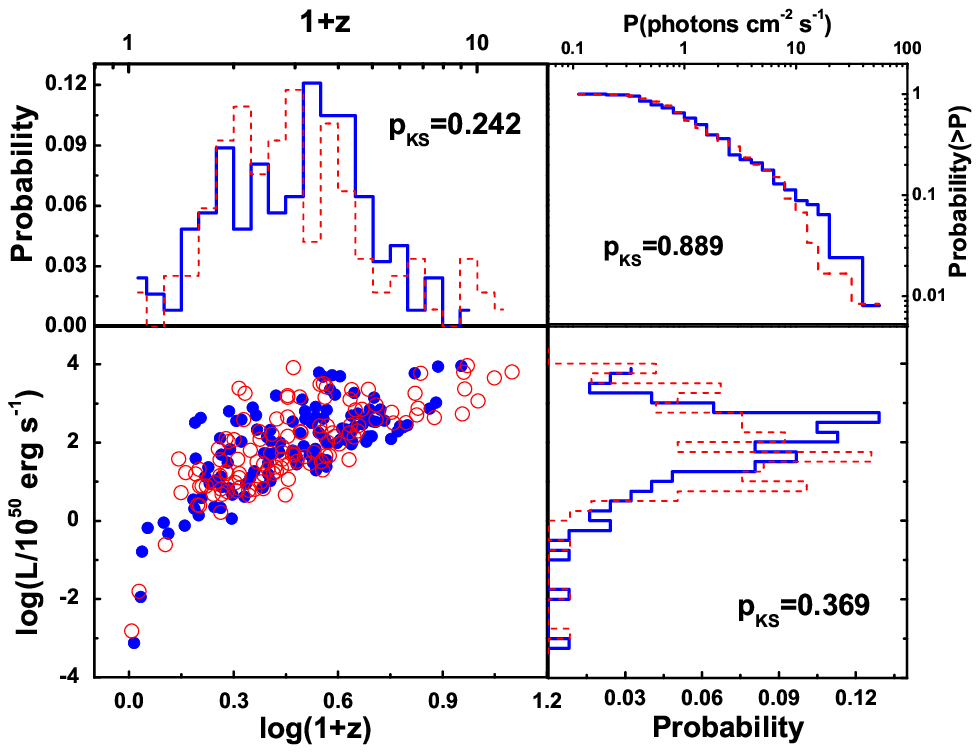}} \caption{The same as Fig. 4, but
for Case C: $R_{\rm LGRB}(z) \propto R_{\rm SFR} (z) (1+z)^{\delta}$.}
\label{Case_C}
\end{figure}

This case assumes that the GRB rate follows the SFR in conjunction with some
sort of GRB rate evolving with redshift characterized by $(1+z)^{\delta}$
(Kistler et al. 2009). We find that $\delta=0.85$ can yield the best
consistency. The results are shown in Fig. \ref{Case_C}. This case well
reproduces the observed $\log N-\log P$, $z$, and $L$ distributions, with
$p_{\rm K-S}=0.889, \ 0.242, \ 0.369$, respectively, for the simulations on the
known-redshift samples, respectively. However, the consistency of the $z$ and
$L$ distributions of the simulations for the triggered sample are only
marginally acceptable (with $p_{\rm K-S}<0.1$).

\subsection{Case D: $R_{\rm LGRB}(z) \propto R_{\rm SFR} (z) f(z)\Theta (\epsilon, z)$}
\begin{figure} 
\resizebox{8cm}{!}{\includegraphics{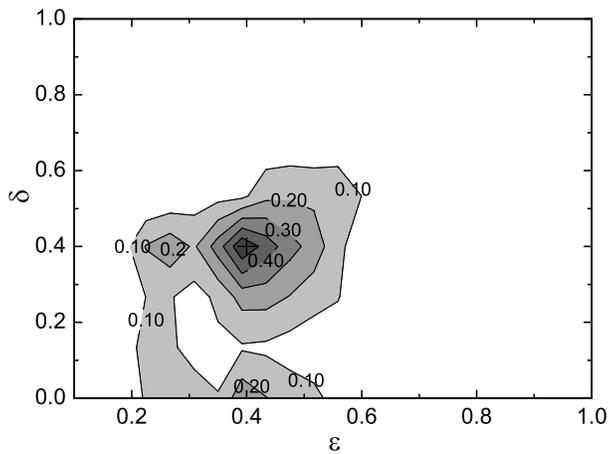}} \caption{Contours of $p_{K-S}$ in
the $\epsilon-\delta$ plane for Case D: $R_{\rm LGRB}(z) \propto R_{\rm SFR}
(z) (1+z)^{\delta}\Theta (\epsilon, z)$, where $p_{K-S}=p^{P}_{K-S}\times
p^{z}_{K-S}\times p^{L}_{K-S}$. The $p^{P}_{K-S}$, $p^{z}_{K-S}$, and
$p^{L}_{K-S}$ are the probabilities of the K-S tests for the $\log N-\log P$,
$z$, and $L$ distributions of the simulated GRB sample with redshift
measurement, respectively. The cross marks the best consistency between our
simulations and observations, i.e., $(\delta, \epsilon)=(0.4,\ 0.4)$.}
\label{Case_D1}
\end{figure}

\begin{figure} 
\resizebox{8cm}{!}{\includegraphics{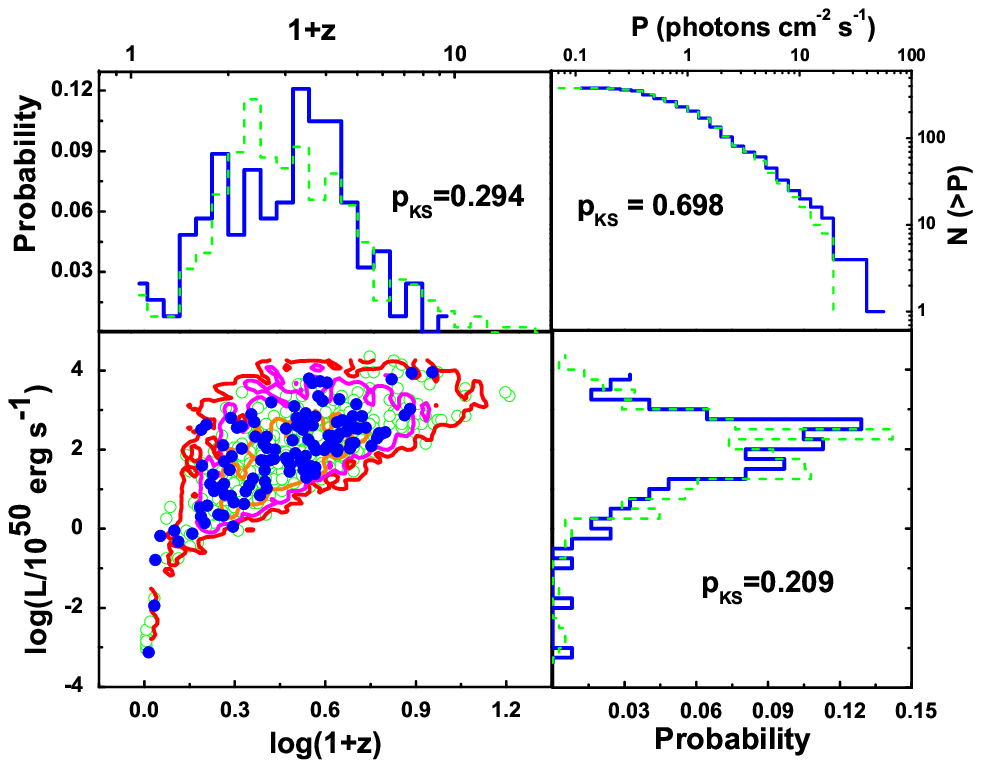}}
\resizebox{8.1cm}{!}{\includegraphics{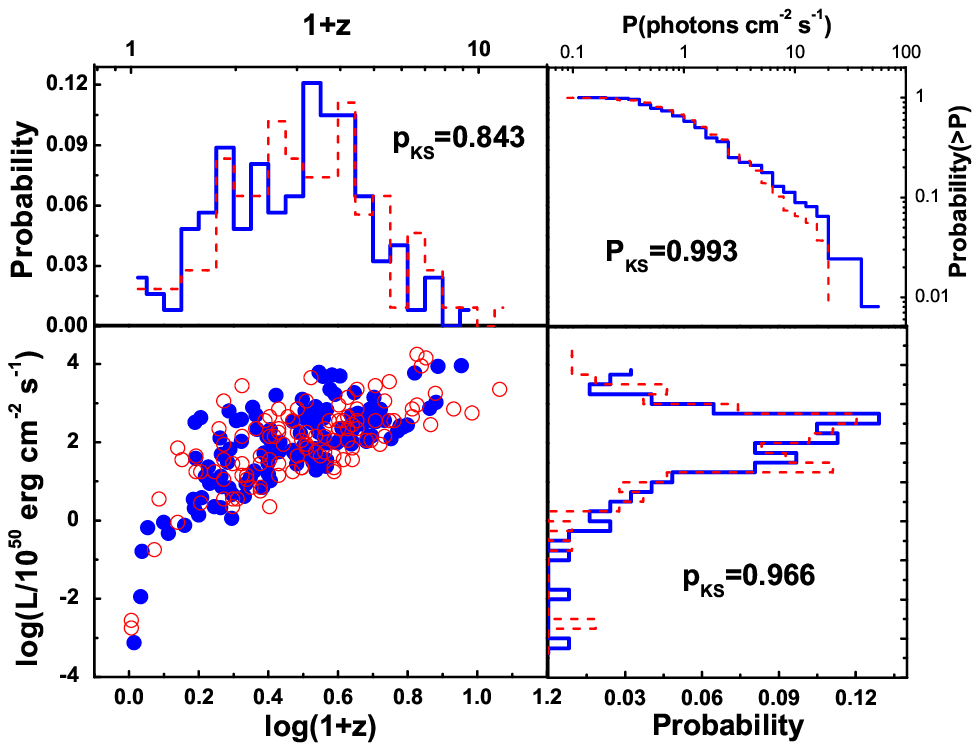}} \caption{The same as Fig. 4, but
for Case D: $R_{\rm LGRB}(z) \propto R_{\rm SFR} (z) (1+z)^{\delta}\Theta
(\epsilon, z)$.} with a parameter set $(\delta, \epsilon)=(0.4,\ 0.4)$.
\label{Case_D2}
\end{figure}

In this case we assume that the GRB rate follows the SFR incorporating with
both GRB rate evolution and cosmic metallicity evolution. We search for the
best $\delta$ and $\epsilon$ by means of a global K-S test probability defined
as $p_{K-S}=p^{P}_{K-S}\times p^{z}_{K-S}\times p^{L}_{K-S}$, where
$p^{P}_{K-S}$, $p^{z}_{K-S}$, and $p^{L}_{K-S}$ are the probabilities of the
K-S tests for the $\log N-\log P$, $z$, and $L$ distributions of the simulated
GRB sample with redshift measurement. The contours of $p_{K-S}\geq 0.1$ are
shown in Fig. \ref{Case_D1}. We find that $0.2<\epsilon<0.6$ and $\delta<0.6$
in case of $p_{K-S}\geq 0.1$. The parameter set ($\epsilon$, $\delta$)=($\sim
0.4$, $\sim 0.4$) produces the best consistency between our simulations and
observations, as show in Fig. \ref{Case_D2}, yielding $p^{P}_{\rm K-S}=0.698$,
$p^{z}_{\rm K-S}=0.294$ and $p^{L}_{\rm K-S}=0.209$ for the triggered sample
and $p^{P}_{\rm K-S}=0.993$, $p^{z}_{\rm K-S}=0.843$ and $p^{L}_{\rm
K-S}=0.966$ for the known-redshift sample.
\section{Discussion}
As shown above, the observed excesses of both high-$z$ GRB rate at $z>4$ and luminous GRBs with $L>10^{53}$ erg
s$^{-1}$ over that predicted by assuming that the GRB rate simply traces the star formation rate (Case A), can
be explained with the cosmic metallicity evolution (Case B) and an unknown GRB rate increase as $(1+z)^{\delta}$
(Case C).

Our simulations of Case D that incorporates Cases B and C well reproduce the
observations. Taking this case as a preferred one, we find that the percentage
of GRBs with $z> 4$ in the current BAT GRB sample is $\sim 15\%$ ($\sim 58$
bursts) and $\sim 4\%$ of the GRBs ($\sim 15$ bursts) have $z> 6$, consistent
with that reported by Butler et al. (2009), but slightly larger than that
predicted by (Daigne, et al. 2006). {\em Swift} has operated for five years. We
thus simulate the observations for a 10 year operation. A uniform sample of
$\sim 1000$ GRBs with $\sim 300$ redshift detections (in the current follow-up
status) would be established in this mission period. Our simulations show that
{\em Swift}/BAT-like instruments can trigger GRBs at redshift up to $14$, and
$\sim 0.5\%$ triggered {\em Swift} GRB may be at redshift $z>10$. Exciting
evidence is from {\em Swift} GRB 100205. The detected high red color of this
GRB in its afterglows, K-H=$1.6\pm 0.5$ mag, likely suggests that its redshift
is $11\leq z\leq 13.5$, if this red color is due to Lyman-$\alpha$ absorption
within the H filter bandpass (Cucchiara et al. 2010). The era of exploring the
deepest Universe with GRBs is coming, e.g., with rapid follow-up and
localization capabilities on the ground and the space-based multi-band
astronomical variable object monitor (SVOM) mission that is being developed in
cooperation between the China and French (G\"{o}tz et al. 2009; Basa, et al.
2008). Thanks to the low energy trigger threshold ($\sim$ 4 keV) of the ECLAIRs
on-board, SVOM is more sensitive than previous missions for the detection of
soft, hence potentially most distant, GRBs. The on-board visible telescope (VT)
would quickly refine the GRB position for three ground based dedicated
instruments, two robotic telescopes (GFTs) and a wide angle optical monitor
(GWAC). Such optimized observing strategy would increase the redshift detection
rate for high-$z$ GRBs.
\section{Conclusions}
Motivated by the increasing detections of high-$z$ GRBs, we have investigated
what may be responsible for the high productivity of GRBs at high-$z$ through
Monte Carlo simulations. We parameterize the trigger probability of {\em
Swift}/BAT using a large GRB sample observed with {\em CRGO}/BATSE. The
redshift measurement probability is also derived from current {\em Swift} GRB
sample. We compare our simulation results to the Swift observations with $\log
N-\log P$ and $L-z$ distributions. We show in the case that the GRB rate is
purely proportional to the SFR, our simulations poorly reproduce the GRBs rate
at $z>4$. We explain this GRB rate excess over that predicted by the SFR with
the cosmic metallicity evolution effect and unknown GRB rate increase as
$(1+z)^{\delta}$. Although the two scenarios can make better consistency
between our simulations and observations, either one cannot simultaneously
reproduce the observations alone. Incorporation of the two scenarios gives
greater agreement between our simulations and observations, indicating that the
combination of both GRB rate evolution and cosmic metallicity evolution would
result in the observed high-$z$ GRB rate excess over that predicted from the
star formation history. We get $0.2<\epsilon<0.6$ and $\delta<0.6$ in case of
$p_{K-S}\geq 0.1$. The parameter set ($\epsilon$, $\delta$)=($\sim 0.4$, $\sim
0.4$) produces the best consistency between our simulations and observations.
Our simulations show that the percentage of GRBs with $z> 4$ in the current BAT
GRB sample is $\sim 15\%$, i.e., $\sim 58$ bursts, and a few triggered LGRBs
may be at $z\simeq 14$. This is sufficient to make GRBs as promising probe for
the high-$z$ Universe.

\section*{Acknowledgments}
We appreciate valuable comments from the anonymous referee. We also thank Bing
Zhang and Fransisco Virgili for helpful discussion. This work was supported by
the National Natural Science Foundation of China under grants (No. 10873002,
10747001, 10821061, 10725313, and 10847003), and the National Basic Research
Program (``973'' Program) of China (Grant 2009CB824800), Guangxi SHI-BAI-QIAN
project (Grant 2007201), the program for 100 Young and Middle-aged Disciplinary
Leaders in Guangxi Higher Education Institution, and the research foundation of
Guangxi University.

\newpage
\begin{table*}
 \centering
 \begin{minipage}{170mm}
  \caption{The selected model parameters from our simulations and the corresponding K-S test probability.}
  \begin{tabular}{@{}cccccccccc@{}}
  \hline
  Case      &   \multicolumn{4}{c}{Model parameters$^{a}$}&& \multicolumn{3}{c}{K-S test$^b$} \\
  \hline
 & $\epsilon$ & $\delta$ & $\rho_{0}^{\rm LL}/\rho_{0}^{\rm HL}$ & $\alpha_{2}^{\rm HL}$& &$p^{P}_{\rm K-S}$ & $p^{z}_{\rm K-S}$  & $p^L_{\rm K-S}$\\
 \hline
 A & --& -- & 50 & 2.5  & &$0.423\mid0.585$& $<0.001$ $\mid$ $<0.001$& $<0.001$ $\mid$ $0.005$\\
 B & 0.5 & -- & 50  & 2.2  & &$0.372\mid0.449$& 0.033 $\mid$ 0.009 & 0.052 $\mid$ 0.292\\
 C & -- & 0.85 & 300  & 2.0 &  &$0.258\mid0.889$& 0.060 $\mid$ 0.242 & 0.018 $\mid$ 0.369\\
 D & 0.4& 0.4 & 200   & 2.0 &  &$0.698\mid0.993$& 0.294 $\mid$ 0.843 & 0.209 $\mid$ 0.966\\
\hline
\end{tabular}

Note: $^a$ The parameters of the luminosity function of LL-GRBs are fixed at
$\alpha_1^{LL}=0$ and $\alpha_2^{\rm LL}=3.0$, $L_b^{\rm LL}=7.5\times10^{46}$
erg s$^{-1}$. For HL-GRBs, $L_b^{\rm HL}$ also fixed at $2.75\times10^{52}$ erg
s$^{-1}$ and $\alpha_1^{\rm HL}$ fixed at 1.36. $^b$ From {\em left} to {\em
right}: The K-S test probabilities of $\log N-\log P$, $\log (1+z)$, and $\log
L$ distributions (triggered GRBs $|$ $z$-known sample).
\end{minipage}
\end{table*}

\newpage


\end{document}